\documentclass[12pt]{article}

\usepackage{graphicx}
\usepackage{scicite}

\usepackage{times}

\def\farcs{\hbox{$.\!\!^{\prime\prime}$}}
\def\fs{\hbox{$.\!\!^{\rm s}$}}
\def\arcmin{\hbox{$^\prime$}}
\def\arcsec{\hbox{$^{\prime\prime}$}}
\def\gapp{\ifmmode\stackrel{>}{_{\sim}}\else$\stackrel{>}{_{\sim}}$\fi}
\def\lapp{\ifmmode\stackrel{<}{_{\sim}}\else$\stackrel{>}{_{\sim}}$\fi}

\newenvironment{sciabstract}{%
\begin{quote} \bf}
{\end{quote}}

\newcounter{lastnote}
\newenvironment{scilastnote}{%
\setcounter{lastnote}{\value{enumiv}}%
\addtocounter{lastnote}{+1}%
\begin{list}%
{\arabic{lastnote}.}
{\setlength{\leftmargin}{.22in}}
{\setlength{\labelsep}{.5em}}}
{\end{list}}

\title{An X-ray Nebula Associated with the Millisecond Pulsar B1957+20}

\author
{B. W. Stappers,$^{1,2\ast}$ B. M. Gaensler,$^{3}$ V. M. Kaspi,$^{4,5}$\\
 M. van  der Klis,$^{2}$ W.~H.~G.~Lewin,$^{5}$ \\
\\
\footnotesize{$^{1}$Stichting ASTRON, 7990 Dwingeloo, The Netherlands}\\
\footnotesize{$^{2}$Sterrenkundig Instituut ``Anton Pannekoek'',1098 SJ Amsterdam, The Netherlands}\\
\footnotesize{$^{3}$Harvard-Smithsonian Center for Astrophysics,60 Garden St, Cambridge, Massachusetts, USA}\\
\footnotesize{$^{4}$Physics Department, McGill University,3600 University Street, Montreal, Quebec, Canada}\\
\footnotesize{$^{5}$Physics Department and Center for Space Research,Massachusetts Institute of Technology, 70 Vasser St.,}\\
\footnotesize{Cambridge, Massachusetts, USA}\\
\\
\footnotesize{$^\ast$To whom correspondence should be addressed; E-mail:  stappers@astron.nl}
}

\date{}

\begin{document} 


\baselineskip24pt


\maketitle


\begin{sciabstract}
 We have detected an x-ray nebula around the binary millisecond pulsar
B1957+20. A narrow tail, corresponding to the shocked pulsar wind, is seen
interior to the known H$\alpha$ bow shock and proves the long-held
assumption that the rotational energy of millisecond pulsars is dissipated
through relativistic winds. Unresolved x-ray emission likely represents the
shock where the winds of the pulsar and its companion collide. This emission
indicates that the efficiency with which relativistic particles are
accelerated in the post shock flow is similar to that for young pulsars,
despite the shock proximity and much weaker surface magnetic field of this
millisecond pulsar.
\end{sciabstract}


Millisecond pulsars are old neutron stars (typically $\sim$ 3\,Gyr) that
have been spun up to a rapid rotation rate ($\lapp 25$\,ms) by accretion
of material from a binary companion\cite{acrs82,bv91}. After the accretion
phase, they appear as radio pulsars with surface magnetic field strengths of
$\sim 10^8$\,G, which, combined with their older ages and rapid rotation rates,
means that they form a separate population from younger pulsars.

PSR B1957+20 is the second fastest-spinning pulsar known, with a rotation
period\cite{fst88} of 1.6\,ms and rotational spin-down
luminosity\cite{tsb+99} $\dot{E}= 1\times10^{35}$\,ergs\,s$^{-1}$. The
pulsar is in a 9.16-hour binary orbit with a low-mass companion star. The
wind of the companion star eclipses the radio emission for $\sim$10\% of
every orbit. The PSR B1957+20 binary system provides an excellent
opportunity to study the wind of a weakly magnetized, recycled neutron
star. The wind is ablating, and may eventually evaporate, the low-mass
companion star. Ablation and heating of the companion star\cite{fbb95} are
believed to be caused by x- or $\gamma$-rays generated in an intrabinary
shock between the pulsar wind and that of the companion star.  Meanwhile, the
high space velocity of the pulsar ($> 220$\,km\,s$^{-1}$)\cite{aft94}, as it
moves through the interstellar medium (ISM), generates sufficient
ram-pressure to confine the pulsar wind and results in the formation of a
bow shock. Upstream of this bow shock H$\alpha$ emission is generated where
neutral material is swept up and collisionally excited; this H$\alpha$
nebula absorbs 1--10\% of the spin-down energy\cite{kh88}. A reverse or
termination shock decelerates the pulsar wind and is located between the
pulsar and the bow shock.

The intrabinary and termination shocks both provide diagnostics of the
pulsar wind. Observations of the interaction of young pulsar winds and the
ISM have shown that they power x-ray emitting synchrotron nebulae which are
typified by the Crab nebula\cite{kc84a}. High-resolution observations of
such pulsar-wind nebulae have shown that the composition of the wind,
characterized by the ratio of the Poynting flux to particle energy flux
$\sigma$, is kinetic energy dominated ($\sigma \approx$ 0.003). However, for
the evolutionarily distinct millisecond pulsar population, the composition,
efficiency, and geometry of the pulsar wind remain unknown. {\it ROSAT}
observations of PSR B1957+20 provided the only previous constraint, and they
have been interpreted as indicating that its wind is different from that of
the Crab pulsar\cite{kpeh92}. However, these data lacked the spatial
resolution and sensitivity to determine the cause of the emission.

We have undertaken a 43\,kilosecond\cite{obs-dur} observation of the millisecond
pulsar B1957+20 using the {\it Chandra} X-ray Observatory. The brightest
x-ray source in the field (Fig. 1) is coincident with the pulsar position. A
tail of x-ray emission is seen extending from the pulsar to the northeast
by at least 16\hbox{$^{\prime\prime}$} with a position angle opposite to the
pulsar's proper motion direction of 212$^{\circ}$\cite{aft94}.

Striking confirmation of the association with the pulsar and the ISM shock
comes by comparing this x-ray tail with the H$\alpha$ bow-shock nebula
(Fig. 2). In bow-shock nebulae, material at the termination shock is
swept back by the ram pressure and forms a cylindrical tube aligned with the
proper motion direction and interior to the bow shock\cite{wlb93,buc02}. The
morphology, direction and location of the x-ray nebula (Fig. 2)
therefore indicate that it corresponds to the termination shock and, in
combination with the enclosing H$\alpha$ emission, demonstrates the expected
double-shock nature of the pulsar's interaction with the ambient medium.

Young pulsar winds are thought to be relativistic, and correspondingly
generate non-thermal emission by synchrotron and/or inverse Compton
processes\cite{aro02,dh92}. However, nothing is known about millisecond
pulsar winds except that the scale of the observed H$\alpha$ nebulae are at
least consistent with their having relativistic winds\cite{cc02}. A
different wind might be expected because of the reduction and possible
alteration of the nature of the magnetic field during the accretion
phase. We must therefore also consider the possibility that they drive
slower winds like that of the Sun, resulting in shock-heated thermal
x-rays. If the spin-down luminosity of a millisecond pulsar is carried away
as kinetic energy in an outflow then $\dot{E} = 0.5\dot{M}V_{w}^2$, where
$\dot{M}$ is the mass loss rate and $V_w$ is the wind speed. Because the
pulsar remains and is at least 1\,Gyr old we know that $\dot{M} <
10^{17}$\,g\,s$^{-1}$, and thus $V_w > 10^{9}$\,cm\,s$^{-1}$.  The
temperature that we measure ($kT \approx 1$\,keV, where $k$ is the Boltzmann
constant) by fitting a spectrum expected for shock-heated gas to the energy
distribution of the 82 counts recorded in the tail region is inconsistent
with such high velocities\cite{spec-warn}.

The emission must therefore be from some form of non thermal process, either
synchrotron or inverse Compton emission, as seen in wind nebulae around
young pulsars. In either case, the x-ray emission requires a population of
relativistic particles in the pulsar wind. Thus, the detection of a distinct
x-ray tail provides direct evidence that millisecond pulsars lose their
rotational energy through relativistic winds.

It is highly unlikely that diffuse shock acceleration\cite{be87} is the
mechanism that generates the emitting particle population in this nebula,
because it is difficult to accelerate particles in a relativistic flow
through this mechanism\cite{at94}.  As in other pulsar wind nebulae, the
acceleration presumably takes place through some other mechanism possibly
involving the role of heavy ions\cite{hagl92}

Because the pulsar and the companion are separated by only
$1.5\times10^{11}$~cm, the intrabinary shock, formed where the pulsar and
companion winds interact, will be located in a strong magnetic field (much
stronger than that at the shock in the Crab nebula).  This intrabinary shock is
therefore a potential source of unresolved synchrotron emission at the
location of the pulsar\cite{at93} and can be used to determine the wind
characteristics at the shock front. The x-ray luminosity in the shock is
dependent on both the post shock magnetic field strength and
$\sigma$\cite{kc84a}. We therefore consider two possibilities for the wind
composition of PSR B1957+20: either dominated by kinetic energy ($\sigma =
0.003$ as seen in the Crab nebula) or magnetically dominated ($\sigma \gg
1$).  We can describe the process by which the spin-down energy of the
pulsar is converted into x-ray emission at the intrabinary shock
by\cite{kpeh92}:
\begin{equation}
f_b \Delta\varepsilon L_{\varepsilon} = f_{rad}f_{\gamma}f_{geom}\dot{E}
\end{equation}
where $L_{\varepsilon}$ is the spectral x-ray luminosity in a band
of width $\Delta\varepsilon \approx$~1~keV centered
on a photon energy $\varepsilon = 1$\,keV. The geometric
factor $f_{geom}$ defines how much of the wind interacts with the
companion, $f_{\gamma}$ is the fraction of the intercepted spin-down
energy flux that goes into accelerating electrons with Lorentz factor
$\gamma$ corresponding to $\varepsilon = 1$\,keV, and $f_{rad}$ is the
radiative efficiency of the corresponding synchrotron emission.  The
fraction of the unresolved x-ray emission that is due to synchrotron
emission produced at the intrabinary shock is $f_b$, the remainder
presumably being emission produced by the pulsar itself.

\begin{table}[hbt]
\caption{Properties of the PSR B1957+20 system\cite{aft94}, and comparison
with the Crab nebula and its central pulsar\cite{tml93}. The fraction of the
wind of PSR B1957+20 that interacts with the companion star wind is
$f_{geom} \gapp r_e^2/(4a^2)\approx 0.02$, where $a$ is the orbital
separation. This fraction is a lower limit, because the pulsar wind is most
likely focused into the equatorial plane\protect\cite{hss+95,mic94} which is
also probably the orbital plane of the binary system\protect\cite{bv91}. The
companion star would therefore intercept more of the spin-down energy from the
pulsar than if it has a spherical wind.}
\vspace{1cm}
\begin{center}
\begin{tabular}{lccc} \hline
Parameter & \multicolumn{2}{c}{B1957+20} & Crab \\
          & ($\sigma = 0.003$)  & ($\sigma \gg 1$) & Nebula / pulsar\\ \hline
Spin period (ms) & \multicolumn{2}{c}{1.6} & 33.5 \\
Surface magnetic field ($10^8$\,G) & \multicolumn{2}{c}{1.4} & $3.8\times10^4$ \\
Distance (kpc) & \multicolumn{2}{c}{1.5} & 2 \\
Age (yr) &  \multicolumn{2}{c}{$>2\times10^9$} & 948 \\ 
Distance to shock (cm) & \multicolumn{2}{c}{$\sim1.5\times10^{11}$} &
          $\sim3\times10^{17}$ \\
\hline
$\dot{E}$ ($10^{35}$~erg~s$^{-1}$) & \multicolumn{2}{c}{1.0} & $4.4\times10^3$\\
$f_b$ & \multicolumn{2}{c}{$\sim0.5$} & $\ldots$ \\
$f_{geom}$ & \multicolumn{2}{c}{$\gapp0.02$} & $\ldots$ \\
$B$ (G) &  2 & 12 & $\approx10^{-4}$ \\
$t_{flow}$ (s) & 5 & 1.7 & $\sim3\times10^{10}$ \\
$t_{rad}$ (s) & 700 & 45 & $\sim5\times10^{11}$ \\
$f_{rad}$ & 0.007 & 0.04 & 0.06 \\
$\varepsilon$ ($\gamma \approx 10^5$) & \multicolumn{2}{c}{1~keV} & 70~$\mu$m \\
$L_\varepsilon$ ($\gamma \approx 10^5$) (erg~s$^{-1}$~keV$^{-1}$) &
\multicolumn{2}{c}{$2.6\times10^{30}$} & $5.8\times10^{40}$ \\
$f_\gamma$ ($\gamma \approx 10^5$) & $<0.09$ & $<0.02$ & 0.04 \\ \hline
\end{tabular}
\label{tab_params}
\end{center}
\end{table}

The flow immediately downstream of the intrabinary shock is expected to
undergo Doppler boosting as it passes around the companion\cite{at93}.
Thus, one expects the x-ray emission at orbital phases before and after
eclipse to be enhanced by up to a factor 2.2 depending on the flow speed and
the degree of absorption and/or scattering by the companion
wind\cite{at93}. In contrast, the x-ray emission at the eclipse (orbital
phase 0.25) may be reduced because of obscuration of the shock by the
companion star. If we can measure these variations then we can determine
$f_b$. The lowest and highest count rates in the folded light curve (Fig 3.)
are during and immediately after eclipse, respectively, with comparatively
low probability of this variation being by chance (see caption of Fig 3.).  If this
apparent orbital modulation is genuine and corresponds to a modulation of
the x-ray flux from the intrabinary shock by a factor of 2.2, then we find that
$f_b \sim 0.5$.

The post shock magnetic field strengths\cite{kc84a} corresponding to our two
limiting values of $\sigma$ ($0.003$ and $\gg1$) are listed in
Table~\ref{tab_params}; the corresponding Lorentz factors of the synchrotron
emitting relativistic electrons for both cases are $\gamma =
2.4\times10^5(\varepsilon/B)^{1/2}\approx 10^5$ at $\varepsilon = 1$\,keV,
where $B$ is the post shock magnetic field strength.  For each case the
emitting region is taken to be the radius of the radio-eclipse
region\cite{aft94} $r_e = 5\times10^{10}$\,cm; for $\sigma = 0.003$ the flow
speed is $v_{flow} = c/3$, whereas for $\sigma \gg 1$ we expect $v_{flow} =
c$\cite{kc84a}. The corresponding residence times are given by $t_{flow} =
v_{flow}/r_e$ (Table~\ref{tab_params}), and the radiative lifetimes of the
emitting electrons are $t_{rad} = 5.1\times10^8/(\gamma B^2)$.  We can thus
compute the radiative efficiency\cite{gs65} in each case as $f_{rad} =
(1+t_{rad}/t_{flow})^{-1}$.  With $f_{geom} > 0.02$ (Table~\ref{tab_params})
and $f_b \sim 0.5$, we can infer that $f_\gamma < 0.09$ ($\sigma = 0.003$) or
$f_\gamma < 0.02$ ($\sigma \gg 1$) for Lorentz factor $\gamma \sim 10^5$.

We directly compared the properties of the shocked wind of PSR B1957+20 with
those derived for the Crab nebula with the same Lorentz factor and find that
$f_{\gamma} \approx 0.04$ for electrons with $\gamma = 10^5$ in the Crab,
which, regardless of the assumed value of $\sigma$, is similar to that which
we derived for PSR~B1957+20. The available evidence therefore suggests that
despite being subject to a prolonged evolutionary process that has altered
its magnetic field by many orders of magnitude and having a shock that
occurs much closer to the pulsar, this millisecond pulsar generates a wind
for which the efficiency of relativistic particle acceleration in this
post shock flow is similar to that seen for the winds of young pulsars. New
{\em Chandra}\ data on this and other pulsars\cite{gak+02,wht+00,hgh01} are
providing the first detailed observational input into studies of
relativistic flows, particle acceleration, and magnetohydrodynamic shocks.



\begin{scilastnote}
\item We thank Jon Arons for useful discussions and Heath Jones for
providing the H$\alpha$ image. This work was supported in part by NASA
through a Chandra X-ray Observatory Guest Observer grant, and by
the Netherlands Organisation for Scientific Research (NWO).
\end{scilastnote}


\begin{figure}
\includegraphics[angle=270,width=12cm]{1079841fig1.eps}
\caption[]{ The full resolution {\it Chandra} image from the ACIS-S3 detector
in the energy range 0.3--10.0 keV. The source at (J2000) RA 19$^{\rm
h}$59$^{\rm m}$36$\fs$75$\pm0\fs$01; Dec
+20$^{\circ}$48$\arcmin$15\farcs0$\pm$0\farcs1 is coincident with the proper
motion-corrected radio timing position of PSR B1957+20, RA 19$^{\rm
h}$59$^{\rm m}$36$\fs$75788$\pm$0\fs00005; Dec
+20$^{\circ}$48\arcmin14\farcs8482$\pm$0\farcs0006\cite{aft94}. Four of the
next brightest x-ray sources in the field are circled and the source labeled
1 is coincident to better than 0\farcs1 with the Tycho II star
1628-01136-1\cite{hfm+00}.  In an aperture of radius 1\farcs5 centered on
the pulsar position we detect a background corrected total of $370\pm20$
counts from the pulsar in the energy range 0.3--10.0~keV. Binning the data
in energy such that each bin contains a minimum of 30 counts, we fit an
absorbed power law resulting in a best fit with hydrogen column density $N_H
= (1.8\pm0.7)\times10^{21}$~cm$^{-2}$, photon index $\Gamma = 1.9\pm0.5$, and
an unabsorbed flux density $F_{x,c} =
6\times10^{-14}$\,ergs\,s$^{-1}$\,cm$^{-2}$ (0.5--7.0 keV) corresponding to
an isotropic x-ray luminosity of $L_{x,c} = 4\pi D^2F_{x,c} =
1.6\times10^{31}D_{\rm 1.5}$ ergs~s$^{-1}$ (0.5--7.0 keV), where $D_{\rm
1.5} = D / (1.5 {\rm kpc})$ is the pulsar distance as derived from its
dispersion measure\cite{tc93}. Counts in the tail region were summed in a
16\arcsec$\times$6\arcsec\ box enclosing the tail and aligned with the
proper motion direction. After background correction, we detect a total of
$82\pm9$ counts from the tail region. Again fitting an absorbed power law
and assuming that the $N_H$ and $\Gamma$ are similar to the values above we
derive an unabsorbed flux density $F_{x,t} =
9\times10^{-15}$\,ergs\,s$^{-1}$\,cm$^{-2}$ (0.5--7.0 keV).}
\label{pure-xray}
\vspace{-5mm}
\end{figure}

\begin{figure}
\includegraphics[angle=0,width=\linewidth]{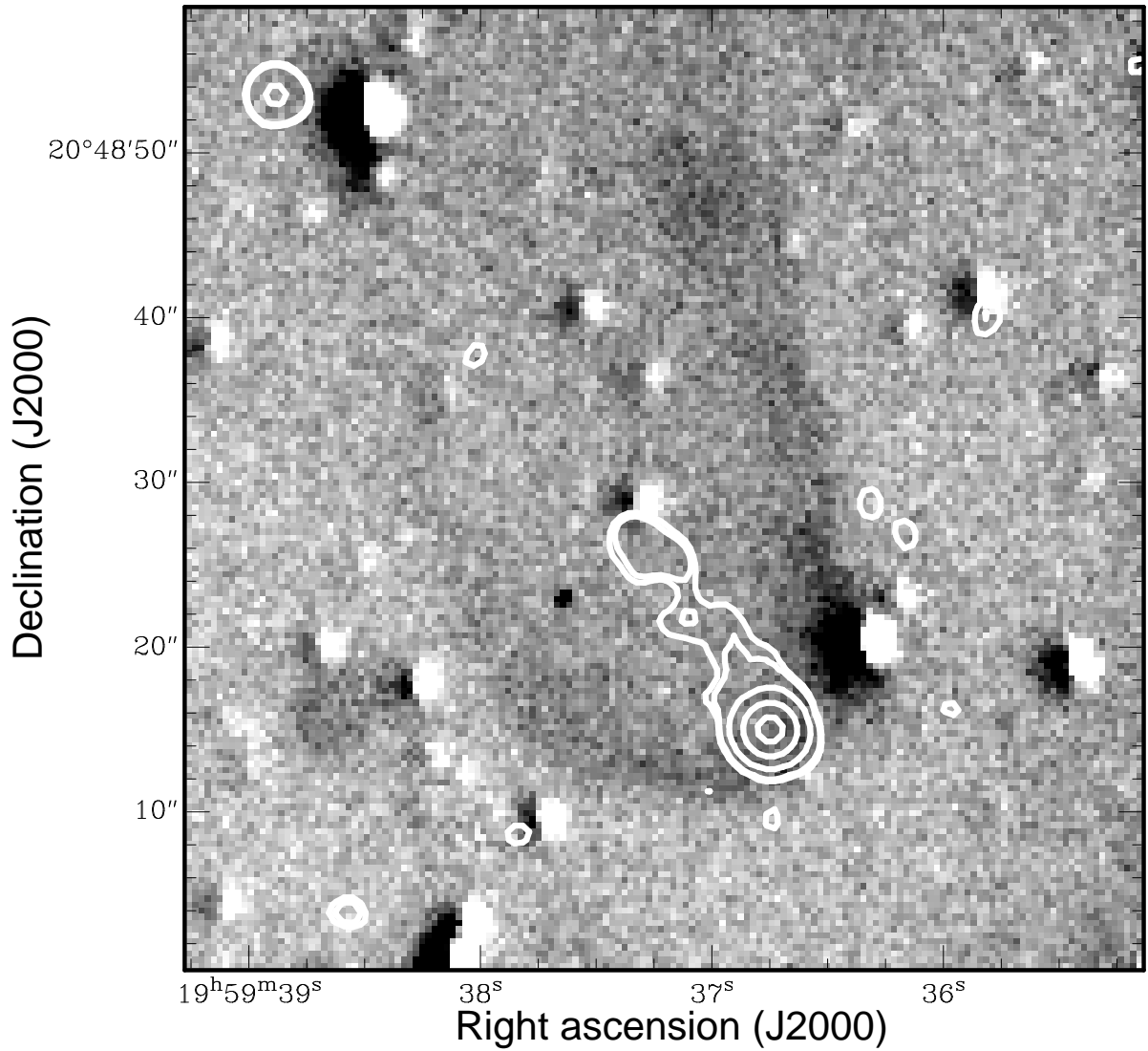}
\caption[] {The same {\it Chandra} data as shown in Figure 1, smoothed to a
resolution of 5\arcsec\ (contours) and overlaid on an H$\alpha$ image
obtained from Taurus Tunable Filter service mode observations on the Anglo
Australian Telescope on 2000 Aug 3. The images were aligned to within
$\sim$0\farcs2 through use of USNO 2.0 stars. The x-ray tail is located well
inside the boundaries of the H$\alpha$ emission and also close to its
symmetry axis. The x-ray contour levels are shown at 0.9, 1.2, 5.3, 35.0 and
78.8\% of the peak x-ray surface brightness. The optical residuals
correspond to incompletely subtracted stars. No optical counterpart was
found to the x-ray source associated with the contours located in the
northeastern corner of the field on a Digitized Sky Survey image.}
\label{ha-xray}
\vspace{-10mm}
\end{figure}

\begin{figure}[hbt]
\includegraphics[angle=270,width=\linewidth]{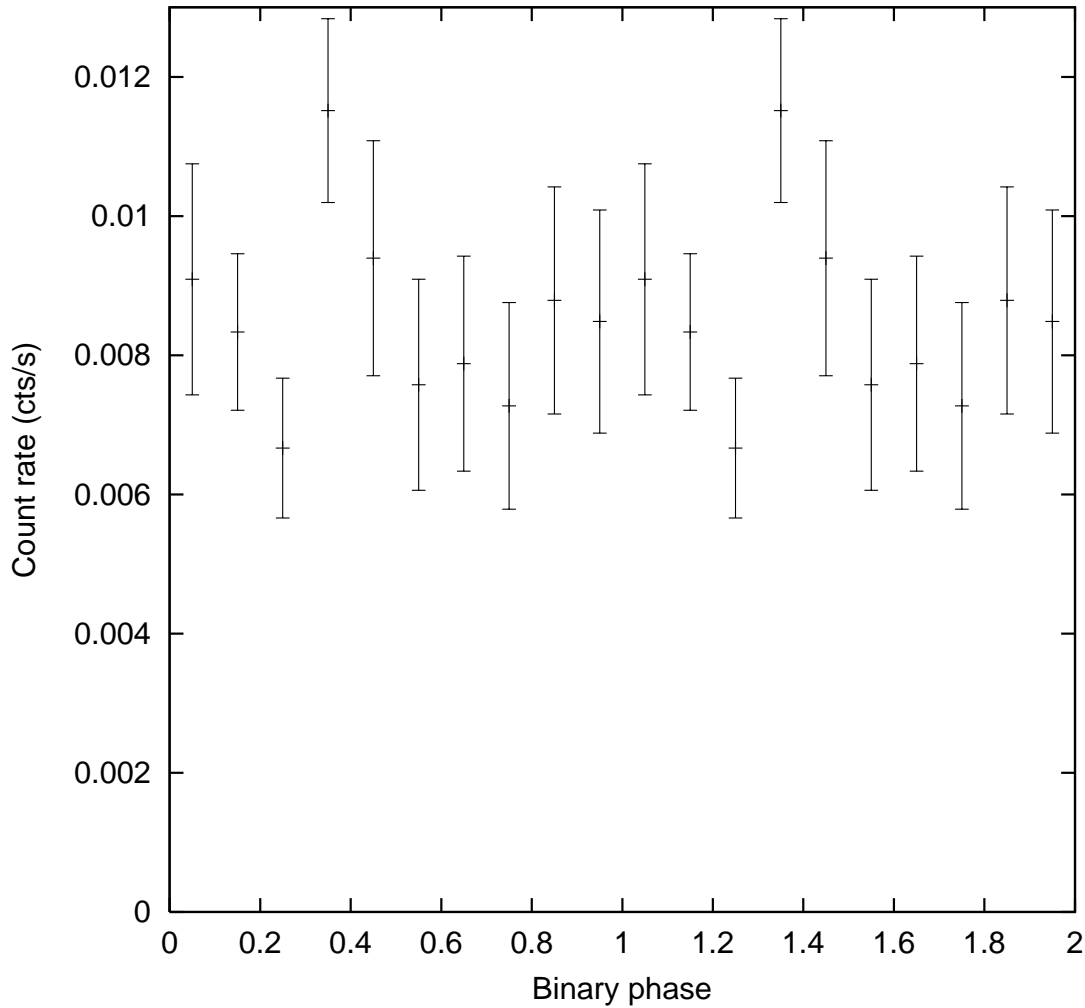}
\caption[] {The light curve of the x-ray emission (0.3--10.0\, keV) from the
point source associated with PSR B1957+20 folded through use of the known
orbital ephemeris (D.\ J.\ Nice, private communication, 2001). The errorbars
correspond to one-sigma Poissonian errors.  Radio eclipse occurs at orbital
phase 0.25. The mean count rate, excluding phases in the range 0.15--0.35
where variations in the x-ray flux are expected, is
(8.4$\pm$0.8)\,cts\,ksec$^{-1}$. Orbital phase bins 0.25 and 0.35 each
correspond to a total observing time of 6600s, and thus we expect to measure
55 counts in each of these phase bins. Using Poisson statistics, the 76
counts detected at phase 0.35 deviate from the expected 55 counts with 99\%
confidence. Scattering and/or absorption of the x-ray emission could reduce
the degree of modulation at either phase 0.15 or 0.35. Taking into account
either possibility, the chance probability for the observed variation at
phase 0.35 is 2\%. A decrease in the count rate is predicted at orbital
phase 0.25 and the 44 counts we detect in this phase bin have a 4\% chance
probability of being drawn from a steady flux distribution.}
\label{lightcurve}
\end{figure}


\begin{thebibliography}{9}
\bibitem{obs-dur} The observation was scheduled so that two eclipses were
observed during the 43\,ksec.
\vspace{-0.5cm}
\bibitem{spec-warn}
We note that this spectrum differs from that used in calculating the flux in
the tail region discussed in the caption to Figure 1.
\vspace{-0.5cm}
\end{thebibliography}
\end{document}